\documentclass[%
reprint,
superscriptaddress,
 amsmath,amssymb,
]{revtex4-1}

\pdfoutput=1

\usepackage{graphicx}
\usepackage{amsmath,latexsym,tabularx}
\usepackage{xcolor}
\usepackage{multirow}
\usepackage[
  colorlinks=true,
  urlcolor=blue,
  linkcolor=blue,
  citecolor=blue
]{hyperref}
\usepackage{multirow,booktabs}
\usepackage{float}
\usepackage{upgreek}
\usepackage{textcomp}

\newcommand{\txsim}{\raisebox{0.5ex}{\texttildelow}}
\newcommand{\s}{{\nobreak\hspace{.2em}}}
\newcommand{\mus}{$\upmu$s}
\newcommand{\mum}{$\upmu$m}
\newcommand{\muW}{$\upmu$W}

\newcommand{\affilLL}[0]{Lincoln Laboratory, Massachusetts Institute of Technology, Lexington, Massachusetts 02421, USA}
\newcommand{\affilMIT}{Massachusetts Institute of Technology, Cambridge, Massachusetts 02139, USA}

\newcommand{\srp}{$^{88}\text{Sr}^{+}$}
\newcommand{\sr}{$\text{Sr}^{+}$}

\newcommand{\specificthanks}[1]{\@fnsymbol{#1}}

\begin{document}

\author{David Reens}
\email[]{david.reens@ll.mit.edu}
\affiliation{\affilLL}

\author{Michael Collins}
\affiliation{\affilLL}

\author{Joseph Ciampi}
\affiliation{\affilLL}

\author{Dave Kharas}
\affiliation{\affilLL}

\author{Brian F. Aull}
\affiliation{\affilLL}

\author{Kevan~Donlon}
\affiliation{\affilLL}

\author{Colin~D.~Bruzewicz}
\affiliation{\affilLL}

\author{Bradley Felton}
\affiliation{\affilLL}

\author{Jules Stuart}
\thanks{Present address: National Institute of Standards and Technology, Boulder, CO}
\affiliation{\affilLL}
\affiliation{\affilMIT}


\author{Robert J. Niffenegger}
\thanks{Present address: University of Massachusetts-Amherst, Amherst, MA}
\affiliation{\affilLL}

\author{Philip~Rich}
\affiliation{\affilLL}
\affiliation{\affilMIT}

\author{Danielle Braje}
\affiliation{\affilLL}

\author{Kevin K. Ryu}
\affiliation{\affilLL}

\author{John~Chiaverini}
\affiliation{\affilLL}
\affiliation{\affilMIT}

\author{Robert McConnell}
\email[]{robert.mcconnell@ll.mit.edu}
\affiliation{\affilLL}

\date{\today}


\title{High-Fidelity Ion State Detection Using Trap-Integrated Avalanche Photodiodes}

\begin{abstract}
Integrated technologies greatly enhance the prospects for practical quantum information processing and sensing devices based on trapped ions. 
High-speed and high-fidelity ion state readout is critical for any such application. 
Integrated detectors offer significant advantages for system portability and can also greatly facilitate parallel operations if a separate detector can be incorporated at each ion-trapping location.
Here we demonstrate ion quantum state detection at room temperature utilizing single-photon avalanche diodes (SPADs) integrated directly into the substrate of silicon ion trapping chips. 
We detect the state of a trapped \sr \s ion via fluorescence collection with the SPAD, achieving $99.92(1)\%$ average fidelity in 450~\mus, opening the door to the application of integrated state detection to quantum computing and sensing utilizing arrays of trapped ions.
\end{abstract}

\maketitle

Trapped ions are a promising technology for scalable quantum information processing \cite{HaffnerIonQC2008, bruzewicz2019trapped} and currently form the basis of the highest-accuracy optical atomic clocks~\cite{Brewer2019}. Integrating elements of these systems may prove essential for scaling quantum information processors, as well as for reducing the size of optical clocks and other quantum sensors for many field-based or space-based applications. Recent advances in integrated control technologies, including photonics for on-chip light delivery~\cite{mehta2016integrated, niffenegger2020integrated, Mehta2020, ivory2021integrated} and electronic control~\cite{stuart2019DACtrap}, highlight the possibility for a dramatic reduction in the hardware overhead for these systems by performing many ion control functions within the substrate of a surface-electrode ion trap. 

Integrating ion detection functions into the trap offers a possibility to reduce system size and complexity by eliminating external lens systems and cameras or photomultiplier tubes typically used for fluorescence-based ion state detection, while also allowing for site-specific readout of a large number of ions. Steps towards this goal have been taken via the use of light-collecting optical fibers~\cite{VanDevenderFiberDetection2010,TakahashiIntegratedFiber2013, ClarkIntegratedOptics2014} or optical cavities~\cite{SterkIonCavity2012} located within the ion-trapping vacuum system, reflective traps \cite{HerskindMirrorTrap2011, vanRynbachMirrorTrap2016}, or chip-integrated optics \cite{MerrillIntegratedOptic2011, JechowIntegratedLens2011, StreedIntegratedOptic2011, Ghadimi2017} for light collection, though in all of these cases a separate photon detector was required outside of the vacuum chamber. A first demonstration of integrated ion detection involved a commercial photodiode attached to a transparent ion trap ~\cite{EltonyITOTrap2013}, although only clouds of ${\sim}50$ ions could be detected, and the 
detector size posed an obstacle to further miniaturization. Recently, a superconducting nanowire single-photon detector (SNSPD) was used to detect the state of a single $^9$Be$^+$ ion in a trap operating at $3.7\s \text{K}$ \cite{todaro2021snspd}, while an integrated avalanche photodiode (APD) detected the presence/absence of a $^{174}$Yb$^+$ ion in 7.7 ms at room temperature \cite{SetzerIntegratedAPD2021}. Nonetheless, rapid, room-temperature, on-chip quantum-state detection of an ion has yet to be achieved. 

We here report the rapid and high-fidelity state detection of \sr \s ions with an efficient, low-dark-current silicon Geiger-mode APD integrated into a surface-electrode ion trap operating at room temperature. In Geiger mode, the APD is biased beyond its breakdown voltage and individual ion fluorescence photons trigger avalanche breakdowns within the APD, generating voltage pulses that can be counted with an external readout circuit; such an APD is referred to as a single-photon avalanche diode (SPAD). A dual-layer indium-tin-oxide (ITO) coating on the trap protects the ion from voltage perturbations due to the pulsing SPAD, allowing stable trapping and coherent operations. We achieve state detection fidelity of $99.92(1)\%$ using a maximum-likelihood estimation technique, in an average detect time of $450$\s\mus\s with an adaptive scheme. The SPAD fabrication is compatible with integrated multi-wavelength photonics processes~\cite{niffenegger2020integrated}, and thus provides a pathway for the elimination of all free-space optics required for ion trapping and control.

\begin{figure}
\centering
\includegraphics[width=\columnwidth]{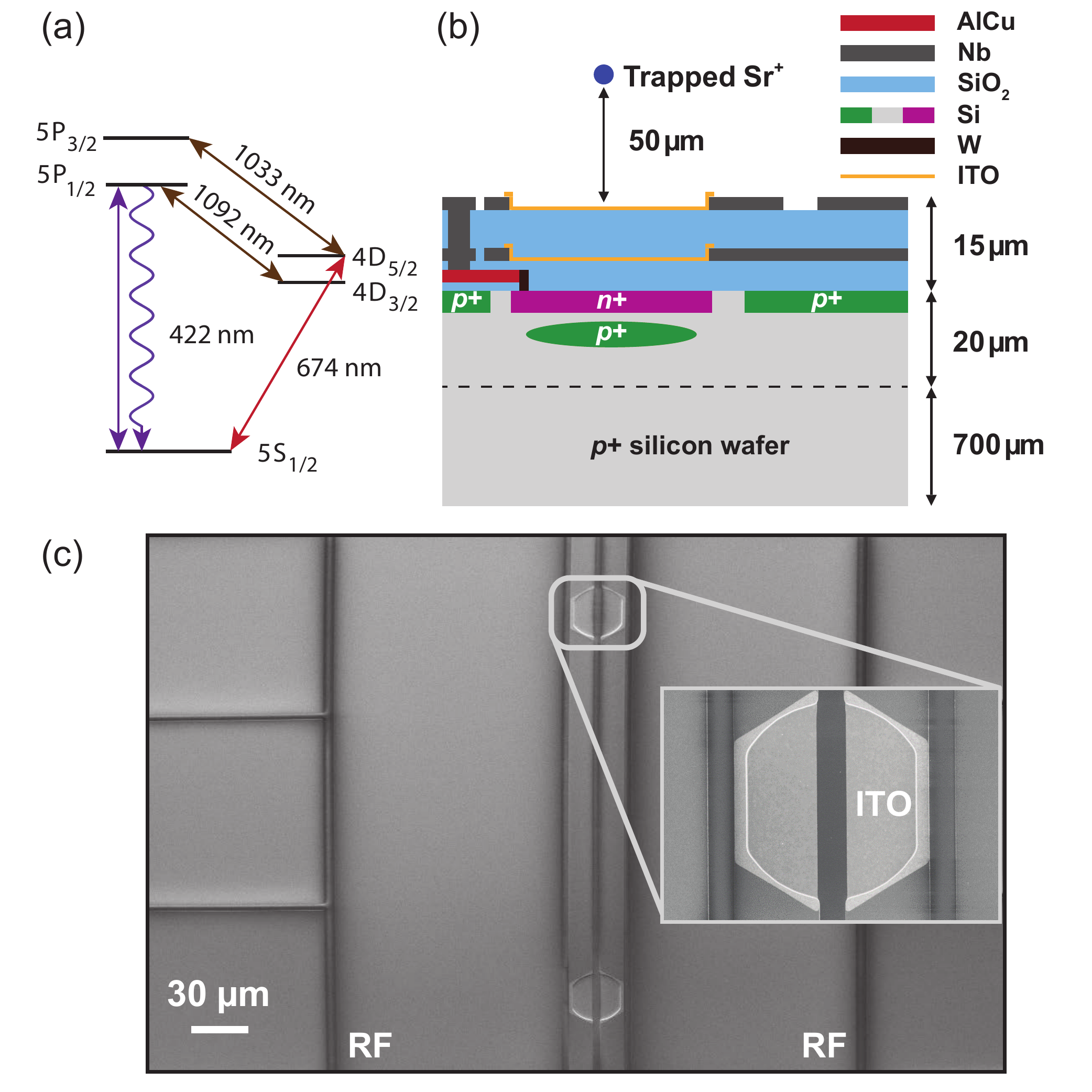}
\vspace{-7mm}
\caption{
\label{fig:fab}
(a) Simplified level structure of the \sr ion. (b) Cross-sectional illustration of the fabricated SPADs within the ion trap chip (not to scale). (c) SEM image of representative device after fabrication (plan view). SPADs with ITO coating (seen in inset) are located directly underneath ion trapping sites and spaced 240\s\mum\s apart. RF voltage is applied to the two indicated electrodes, while DC voltages are applied to the remaining electrodes to generate an ion-trapping potential as described in the text.}
\vspace{-3 mm}
\end{figure}

Figure \ref{fig:fab}(a) shows a simplified level structure of the \sr \s ion. We store quantum information in the optical clock/qubit states, the ground $5$S$_{1/2}$ and the metastable $4$D$_{5/2}$ state (lifetime 390 ms \cite{safronova2011excitation}), which can be coherently coupled via a 674-nm laser.
A 422-nm laser is used for Doppler cooling on the broad $5$S$_{1/2}$ $\rightarrow$ \s $5{\textrm P}_{1/2}$ transition. This transition is also used, during readout, to produce fluorescence photons from an ion in the ground state; because the 422-nm laser does not couple to the $4$D$_{5/2}$ state, the latter will appear dark, thus allowing discrimination between the bright $5$S$_{1/2}$ and dark $4$D$_{5/2}$ state.
Additional lasers at 1033~nm and 1092~nm are used for quenching of the metastable state and for pumping out of the undesired $4$D$_{3/2}$ state, respectively.

Figure 1(b) shows a cross-sectional view of the fabri-
cated  ion  traps  with  integrated  SPADs. SPADs are fabricated on a p-type Si wafer and are optimized for performance at 422 nm, but could also be used for detection at other visible wavelengths. Details of SPAD fabrication can be found in the Supplemental Material. After SPADs were fabricated and tested, a Nb metal ground plane was deposited on top of a 2-$\mu$m-thick oxide layer, with apertures to allow electrical connections and optical access to the SPADs. A second, 10-$\mu$m-thick oxide layer was then added, followed by patterned Nb electrode metal to form a linear surface-electrode Paul trap \cite{NIST:SET:QIC:05}. SPADs are located directly under seven ion-trapping sites on the 1~cm $\times$ 1~cm trap. The diameter of each SPAD's photo-sensitive region is $40$~$\mu$m, but ion trap electrodes obscure a portion of the SPAD active area and result in a $30$-$\mu$m-diameter clear aperture (Fig.~\ref{fig:fab}(b)). Two layers of ITO were used, one over the SPAD aperture in the ground plane and one over the similar aperture in the trap metal, to shield the ion from SPAD pulses and to shield the SPAD from trap radio frequency (RF) voltage ($\sim 50$\s V amplitude). A scanning-electron microscope (SEM) image of the finished devices is shown in Fig. \ref{fig:fab}(c).

\begin{figure}[t]
\centering
\includegraphics[scale=0.58]{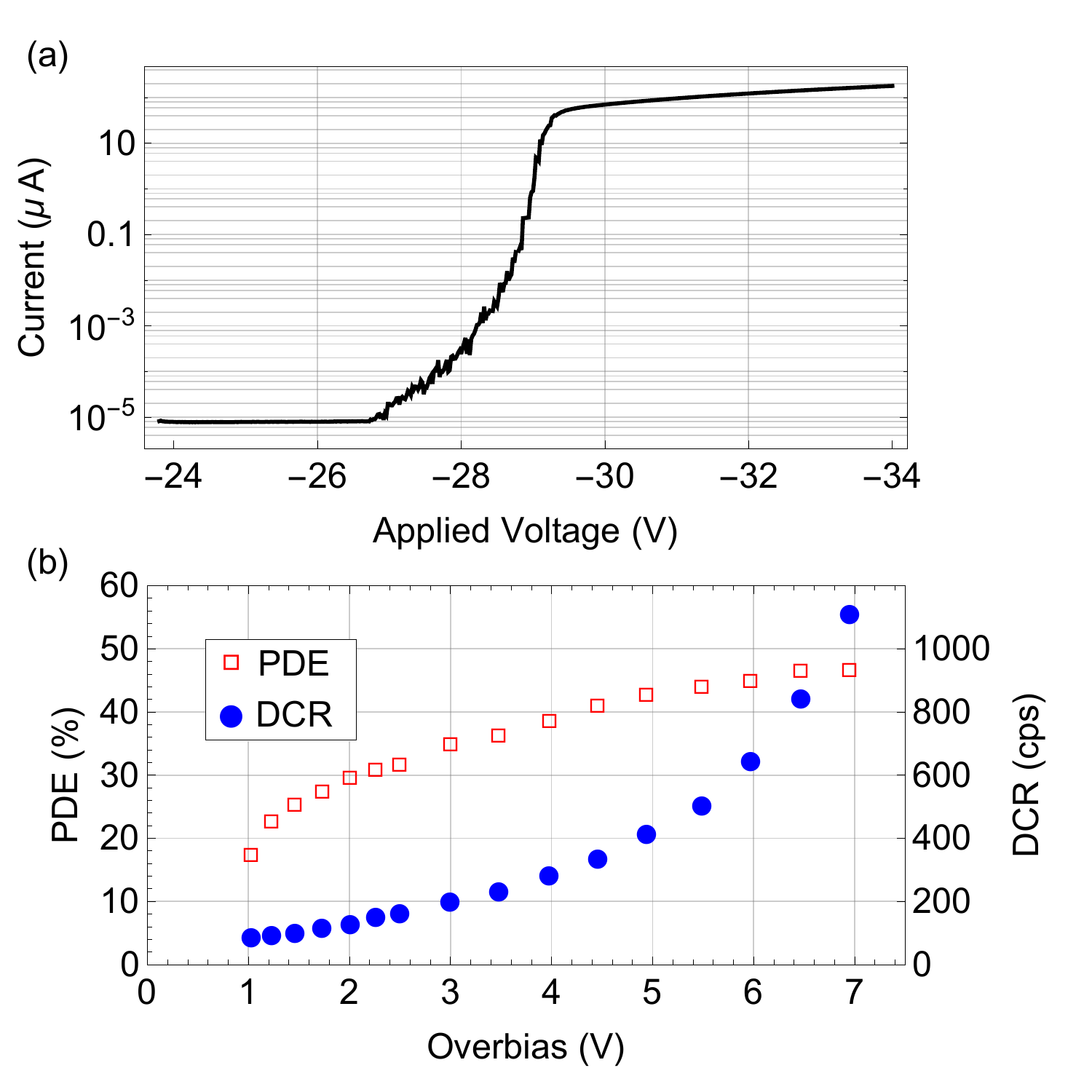}
\vspace{-7 mm}
\caption{
\label{fig:PDE}
Measured SPAD characteristics. (a) Current vs. voltage curve of a representative device, indicating breakdown voltage near $-28$~V; the measurement is taken with a current limit of $200$ $\upmu$A. (b) External photon detection efficiency (PDE, red squares) at wavelength of $420$~nm and device dark count rate (DCR, blue circles) of a representative device as a function of device overbias. Error bars are comparable to the point size.}
\vspace{-3 mm}
\end{figure}

We characterized the SPADs' current-voltage characteristic, external photon detection efficiency (PDE) at the ion fluorescence wavelength of 422~nm, and dark count rate (DCR). The PDE was measured using a time-to-first-pulse method (see Supplemental Material) to avoid overestimating SPAD efficiency due to afterpulsing effects.  Fig.~\ref{fig:PDE} shows measurements of these parameters for a typical SPAD from the same wafer, and of the same design, as the device used for ion-trap measurements.
Panel (a) shows SPAD current versus applied voltage while in darkness, indicating a breakdown voltage of approximately $-28$~V, while panel (b) shows the PDE and DCR versus excess reverse bias relative to this breakdown voltage (overbias). 
The device we use for state detection measurements exhibits a DCR of $109\pm2$~counts per second (cps) at $2$~V overbias, slightly below the value obtained from the measurement in panel (b).

To characterize ion-detection performance of the integrated SPAD trap, we install one of these traps into a room-temperature ion-trapping system, described in the Supplemental Material. Briefly, we trap single \srp \s ions 50\s\mum\s above the trap with typical axial frequency of $2 \pi \times 700$ kHz and radial frequencies of ${\sim} 2 \pi \times 5$~MHz. The experiments described in this work were primarily performed over a single SPAD on the trap, but we observed similar performance when trapping and detecting an ion over a second device on this same trap.

The trap-integrated SPAD is read out using a passive quenching circuit (see Supplemental Material), with a ${\sim}500$~ns quench time and $3.5$~\mus\ recharge time  ($1/e$ times). 
Optimum performance, taking into account speed limitations due to readout-circuit capacitance, is obtained with a $2~$V overbias for a total count rate $106$~kcps ($38$~kcps) while detecting the bright (dark) ion state, with background counts dominated by laser scatter from the $422$-nm detection and $1092$-nm repump beams.

\begin{figure}[t]
\centering
\includegraphics[width=\columnwidth]{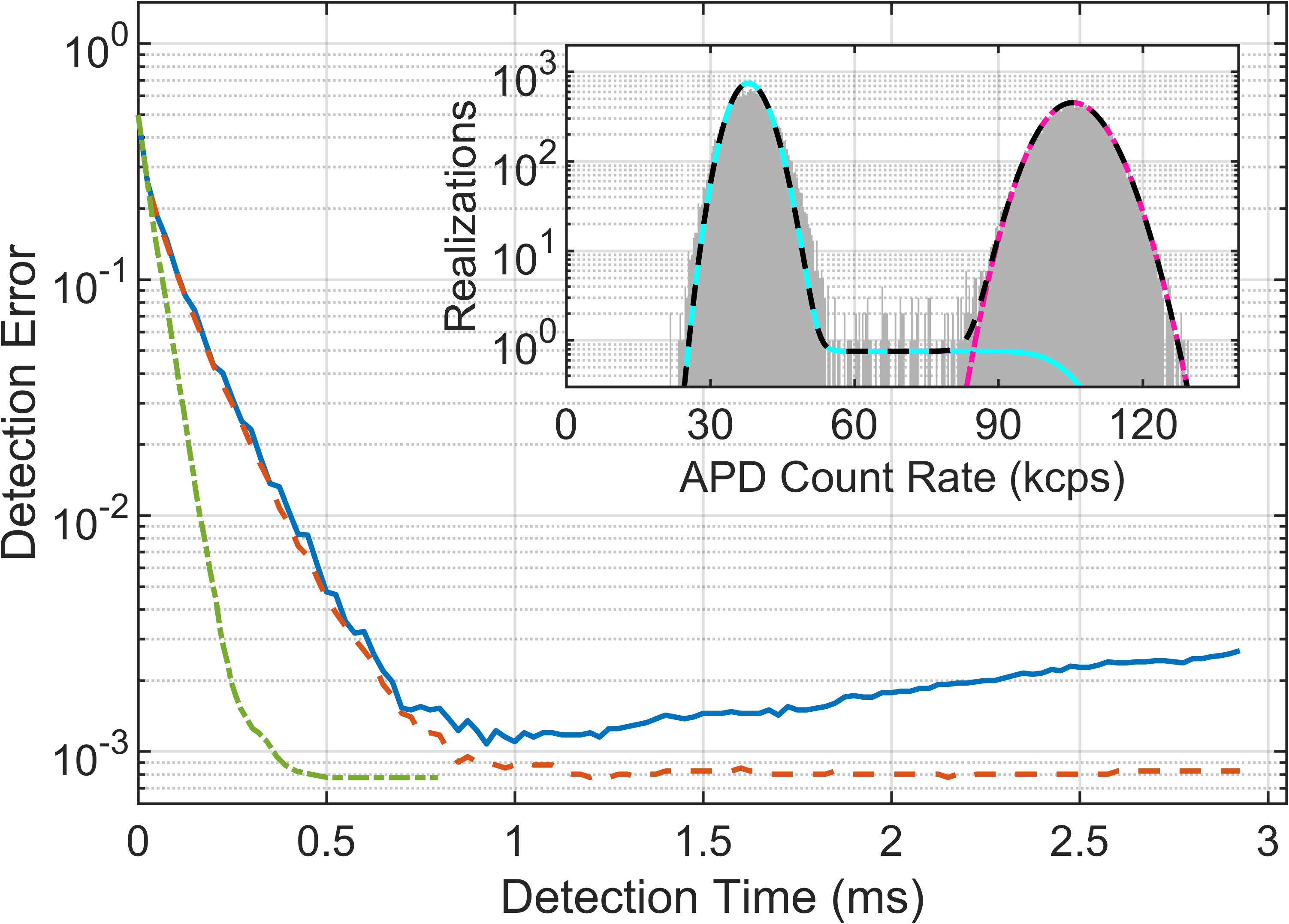}
\vspace{-6 mm}
\caption{
State detection error for simple thresholding (blue), maximum likelihood analysis (red dashed), and using an adaptive readout scheme (green dash dot). 
Inset: Histogram of counts recorded by SPAD in 40,000 ion state-detection trials with detection time of 2.9 ms. 
The fit (black dashed) is the weighted sum of a bright-state Poisson distribution (pink), and a dark-state distribution (cyan) that includes a plateau corresponding to metastable decay during measurement.
The dark state distribution exhibits slightly increased variance relative to the Poissonian expectation, which is accounted for during likelihood estimation.
\label{fig:detect}
}
\vspace{-3 mm}
\end{figure}

We define ion state detection fidelity as $F = 1-(\epsilon_{\textrm{bright}}+\epsilon_{\textrm{dark}})/2$, with $\epsilon_i$ the detection error when the ion is in state $i$. 
To avoid the necessity of high-fidelity dark state preparation, we prepare the ion in a nearly-equal superposition of bright and dark states via a $\pi/2$ pulse on the $5$S$_{1/2}\rightarrow 4$D$_{5/2}$ transition. 
We perform detection simultaneously with the integrated SPAD and with a high NA lens focused onto a photomultiplier tube (PMT) located outside of the vacuum chamber.
As the PMT is far less susceptible to laser scatter due to its distance from the detection beams and to spatial and spectral filtering provided by the high-NA lens and a 422-nm bandpass optical filter, we use it as a reference against which to compare the SPAD.
We determine the SPAD error rates $\epsilon_{\textrm{bright}}$ and $\epsilon_{\textrm{dark}}$ from the discrepancy between SPAD and PMT state identification and average to find the mean infidelity, taking into account the PMT detection error separately as described below. Detection data are binned at $25~\mu$s intervals and post-processed to determine readout error as a function of detection time (see Fig.~\ref{fig:detect}).

Discrimination between the bright and dark states can be performed in multiple ways. 
The most straightforward establishes a threshold of counts in a given time---based on the observed bright and dark count rates from the ion---with the ion being assigned to the bright (dark) state if counts exceed (do not exceed) the threshold. 
Simple thresholding in our system achieves $F = 99.89(1)\%$ in $950$\s\mus\s with a threshold of $66.5$~counts; the prepared superposition is found to project to the bright state in $49.8\%$ of trials. 
A readout method utilizing more information is maximum likelihood estimation~\cite{MyersonAdaptive2008}, which compares the probabilities of the observed count distributions for bright and dark ions via an iterative algorithm. 
Maximum-likelihood estimation deals better with those instances in which the metastable dark state decays during measurement and achieves $F = 99.92(1)\%$ in 1.2~ms in our system. 
Detection speed can be improved with an adaptive readout scheme~\cite{MyersonAdaptive2008}, which is similar to the maximum-likelihood method but concludes measurement when the expected probability of a correct detection exceeds a user-specified threshold. 
In our system adaptive readout also reaches $F=99.92\%$, but in an average detection time of $450$~\mus. 
The time required for adaptive detection is primarily limited by scatter from the detection beams.
We estimate that if stray light could be eliminated, adaptive detection could achieve $F=99.9\%$ in $75$~\mus\s, or $F=99.98\%$ in $125$~\mus, given the characteristics of our SPAD and biasing circuit.

Our fidelity measurements must take into account the possibility of PMT detection errors. The PMT has an expected error rate of $\epsilon_\text{PMT}=3.5\pm0.1\times 10^{-4}$, obtained by simulating the maximum likelihood detection process including metastable decay. The PMT error is dominated by instances when the dark-state ion decays very early during the measurement, such that insufficient time is available to distinguish between the bright and dark ion states. In these instances, simulations suggest that the SPAD is at least $90\%$ likely to also incorrectly detect the state. Because of this correlation of SPAD and PMT errors, we conservatively assume that all PMT errors give rise to SPAD errors. The SPAD detection error ($1-F$) shown in Fig.~\ref{fig:detect} and throughout this work is thus the sum of the measured rate of disagreement between the SPAD and PMT and the estimated PMT error $\epsilon_\text{PMT}$.

\begin{figure}[t]
\centering
\includegraphics[width=\columnwidth]{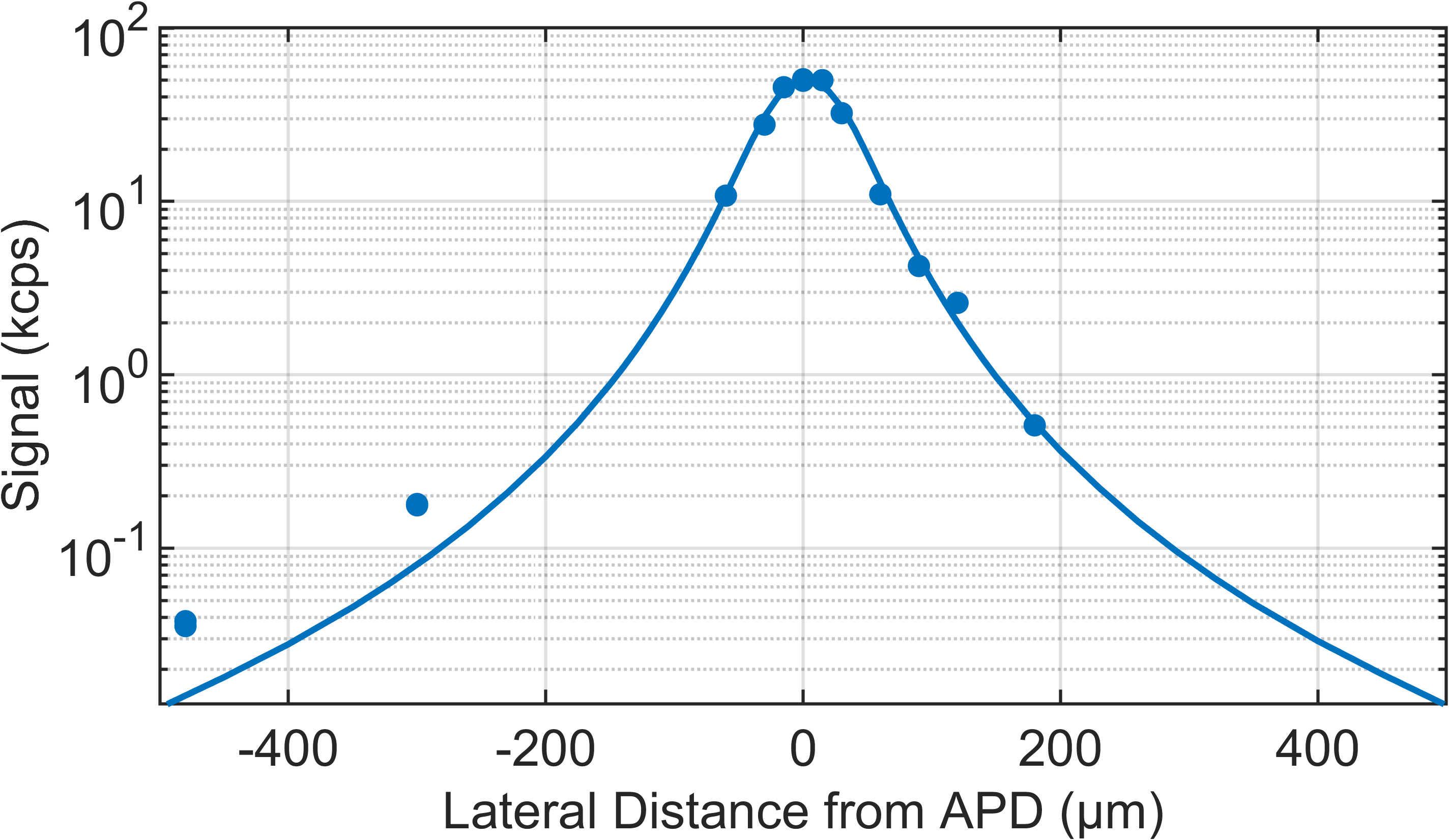}
\vspace{-7 mm}
\caption{
\label{fig:cross}
Measured counts from the SPAD as a function of ion lateral distance from SPAD center. The model line accounts for the full geometry and for reflection and refraction at dielectric interfaces between the trap surface and the SPAD. Statistical error bars are smaller than the size of the markers.} 
\vspace{-3 mm}
\end{figure}

Low optical and electronic crosstalk between different detectors in an on-chip array will be required if they are to be used in larger sensing or information processing ensembles. To characterize potential optical crosstalk in a SPAD array, we measured the dependence of the SPAD count rate as the ion trapping location was translated along the axial direction away from the SPAD.
For these measurements, we prepare a superposition state as described previously, use the PMT to determine the state of the ion, and subtract the average SPAD background count rate from the average bright-ion count rate. 
The count rate for ions directly above the SPAD is 60~kcps, which declines to only 0.18~kcps at a lateral distance of 300~\mum\ and to 0.037~kcps at a distance of 480~\mum.
At smaller values of ion-SPAD distance, the count rate is found to obey a ray-tracing model accounting for full geometry (Fig.~\ref{fig:fab}) including reflections and refractions at the dielectric interfaces between the trap surface and the SPAD.
The discrepancy between model and experiment at larger distances may be attributable to light which passes through gaps between trap electrodes and becomes confined in the oxide layer separating the metal ground plane from the electrodes (Fig~\ref{fig:fab}b), eventually reaching the SPAD. 
Nevertheless, the results suggest that, for ion array site spacings greater than $300$\s\mum, additional background counts due to optical crosstalk would have a negligible effect on detection given the $38$~kcps laser scatter rate. 
Even with stray light eliminated, an additional $200$~cps background only increases the time required for $99.98\%$ detection fidelity from $125$~\mus\s to $140$~\mus.

We additionally perform a measurement to detect electronic crosstalk between SPADs, by trapping an ion above one SPAD, detecting with a SPAD $480$\s\mum\s away, and enabling or disabling the readout circuit for the first SPAD.
Biasing the SPAD below the ion reduces the count rate of the distant SPAD by $40 \pm 20$ cps, an amount small compared to the 68 kcps count rate induced by a bright ion.
This effect, which is likely due to a slight reduction in the common bias voltage shared by the SPADs, does not pose an obstacle to reading out moderate-scale SPAD arrays with this spacing, and could be further mitigated by improving bias stiffness.

\begin{figure}[t]
\centering
\includegraphics[width=\columnwidth]{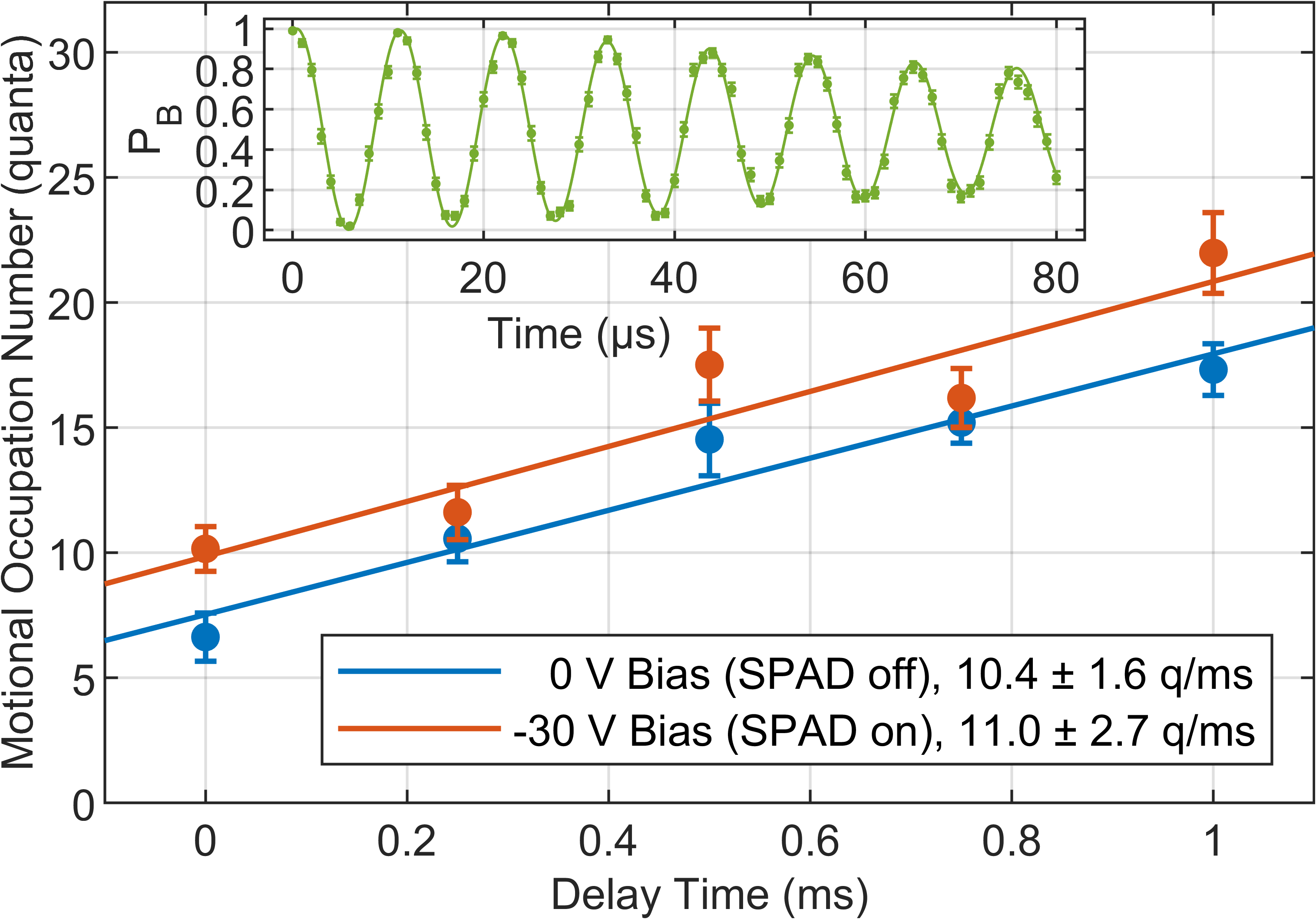}
\vspace{-6 mm}
\caption{\label{fig:heating}
Occupation numbers of a radial motional mode with $\omega = 2\pi\times 5$~MHz are measured after varying delays, with the SPAD either off or on, and data collected with an external PMT. Lines are linear fits with resulting heating rates given in the legend.
Inset: Rabi oscillations on the $5{\textrm S}_{1/2} \rightarrow\ 4{\textrm D}_{5/2}$ transition are fitted to obtain the average motional occupation number for each delay. 
The data and fit shown here correspond to $\bar{n}=6.3\pm0.3$~quanta, and Rabi frequency $\Omega=2\pi\times 94.7\pm0.1$~kHz. P$_{\textrm{B}}$: Bright state probability.}
\vspace{-3 mm}
\end{figure}

In order to be useful for quantum information processing or sensing purposes, integrated detectors must be compatible with high-fidelity quantum control of trapped ions. We demonstrate quantum control of the  ion above the pulsing SPAD by performing Rabi oscillations via the 674-nm beam. Fig.~\ref{fig:heating}(inset) shows a typical Rabi oscillation with $5$\s\mus\s $\pi$-pulse time and fitted contrast of $99.8\%$, limited by the ion's initial Doppler-cooled temperature.

The motional heating rate experienced by ions close to the SPAD is a key figure of merit for  the compatibility of the device with the coherent operations required for atomic clocks or quantum computations. We measure the heating of a radial mode with frequency $\omega = 2 \pi \times 5$ MHz by fitting the decay of contrast seen in Rabi spectroscopy of the metastable-state carrier transition as a function of delay time. We note that, during installation of this trap, some wirebonds to electrodes near the trapped ion's location detached, making full elimination of the ion's excess micromotion difficult. After optimizing the ion's electric-field compensation with the remaining electrodes to minimize the ion's initial average motional excitation $\bar{n}$, we obtain heating rates of approximately $10$~quanta/ms directly above the SPAD (Fig.~\ref{fig:heating}). We find that the observed heating rate does not depend on SPAD bias voltage within our margin of error.

The observed heating rate is high compared to that 
measured at $295\s$K for Nb ion traps of similar electrode configuration~\cite{sedlacek2018}. Because of the difficulty in minimizing micromotion in this trap, it is difficult to unambiguously determine the cause of the higher heating rate in this system. 
We observe rates as low as $0.5$~quanta/ms at a distance $60$~\mum\s away from the SPAD, potentially implicating the SPAD or ITO.
Future investigations may be able to more fully characterize the heating rate over the SPAD, and adapted processing or surface modification, such as surface milling by Ar ions~\cite{HiteIonMill2012,sedlacek2018,HaffnerFirstIonMilling2014}, could be effective in reducing it.

In summary, we have demonstrated rapid and high-fidelity ion state detection with a chip-integrated SPAD, achieving $99.92(1)\%$ detection fidelity in $450$ \mus \s with an adaptive technique. The demonstrated SPAD performance lends itself to on-chip, site-specific detection of arrays of ions for optical clocks or quantum information processing purposes. We also demonstrate low levels of optical and electrical crosstalk between multiple SPADs on the chip separated by hundreds of microns, particularly relevant for optical clocks (where clock ions may be individually trapped and do not necessarily need to be entangled), or for
a quantum computing architecture in which ions are shuttled between different sites for entangling and readout operations~\cite{NIST:ExpIssueswithIons:JresNIST:98,KielpinskiQCCD2002}. 
These SPADs may also be used in on-chip remote-entanglement generation if light collected from two or more ions can be routed through an integrated beamsplitter to SPADs such that ``which-path'' information is removed.

The currently achieved fidelity is limited by scattered light from the measurement and repumping lasers; if eliminated, infidelities and detect times could be reduced four-fold. As more ion-control components---particularly integrated photonics---begin to be incorporated into surface-electrode traps, it will be necessary to carefully screen out any light scattered by on-chip components to prevent overwhelming integrated single-photon detectors. Further improvements in detection time and fidelity might be achieved with an active quenching system, reducing the recovery time of the SPAD at the expense of additional experimental complexity. We briefly note that we have tested these SPADs at cryogenic temperature and found them to operate, but with a roughly two-orders-of-magnitude lower PDE, precluding rapid ion-state detection. Future work may focus on SPADs optimized for cryogenic temperatures and on reducing the observed ion motional heating rate near the SPAD via surface treatments or low-temperature operation.

We thank S. L. Todaro for helpful manuscript comments, M. Purcell-Schuldt for trap layout, and P. Murphy and C. Thoummaraj for chip packaging. This material is based on work supported by the Defense Advanced Research Projects Agency under Air Force Contract No. FA8702-15-D-0001. Any opinions, findings, conclusions or recommendations expressed in this material are those of the author(s) and do not necessarily reflect the views of the Defense Advanced Research Projects Agency.

\section{Supplemental Material}

\subsection{Fabrication}

SPADs are fabricated on a high-resistivity ($>$~$100~\Omega\cdot$cm) p-type silicon layer on a
heavily doped p-type substrate grown via homoepitaxy to enable collection of hole current via anode connection from the entire wafer without additional processing.
Wafers are oxidized to provide passivation and reduce implant channeling effects. 
A shallow boron implant defines a p-type silicon region outside the detector to remove charge generated beyond the device perimeter. 
A shallow arsenic implant is used for the cathode; this step is followed by heavy boron doping to define the regions where anode contacts will be made. 
The wafer is annealed at $1000\s^\circ$C to activate the doping and reduce implant defects. 
Finally, a deep boron implant is performed to define a high-field region for initiating the avalanche process. 
The thermal oxide is then stripped in buffered oxide etch and the silicon is allowed to re-oxidize to provide a low interface-state oxide. 
A 1-$\mu$m-thick oxide is deposited via plasma enhanced chemical vapor deposition, and AlCu contact metal is deposited and patterned and is connected to the SPAD by W vias with Ti/TiN liner. A ground plane of thickness $500$~nm is deposited on top of the first oxide layer, followed by a second oxide layer of thickness $10$~\mum. Linear ion-trap surface electrodes are formed by depositing $1$~$\mu$m of Nb on top of the second oxide layer, with gaps between electrodes formed by standard photolithography and etching techniques. Subsequent to ion-trap fabrication, wafers were sintered in a 100\% hydrogen atmosphere at $400\s^\circ$C for 1~h to reduce the dark-count rate by passivating oxide interface states.

Device simulations were performed to identify optimal implant conditions. 
As the wavelength of interest ($422$~nm) has shallow absorption depth (approximately $200$~nm~\cite{Green1995}), most photons are absorbed at the surface of the silicon. The generated electrons are immediately collected by the cathode, and the holes travel into the multiplier region toward the anode. This process results in the avalanche being primarily initiated by holes rather than electrons~\cite{Oldham1972}. 
The doping of the deep boron implant was increased compared to devices fabricated previously~\cite{AullAPDs2018}, which we observe to result in an increase in the PDE by a factor of approximately $2.5$ at the 422-nm wavelength of interest.

With the SPAD operating voltage near 30 V and typical pulse height of a few volts, there is the potential for ions to be ejected from the trap due to the SPAD pulses. 
It is likewise important to protect the SPAD from the high RF voltage used for ion trapping. 
Simulations in finite-element modeling software indicated that a single 20-nm layer of the transparent conductor indium tin oxide (ITO) in the ground plane provided a shielding factor of approximately 2000.
A second layer of ITO over the aperture in the trap electrodes was added to reduce the effect of possible charging of the exposed dielectric.
These simulations also indicated that RF pickup at the SPAD location is reduced by these dual ITO layers by a factor of $3 \times 10^{5}$.

\subsection{Time-to-first-pulse method for PDE measurement}

The PDE was measured using a time-to-first-pulse method (TTFPM) with a 420-nm laser focused to a spot smaller than the high-gain region of the SPAD. 
The laser emits $70$~ps output pulses whose energy was measured with a NIST-traceable photodiode and then attenuated by $69.8~$dB with neutral-density filters (also calibrated at the operational wavelength). 
The TTFPM avoids overestimating SPAD efficiency due to afterpulses that can arise from charges that become trapped in the device's gain region during a pulse. 
For the TTFPM, we first arm the SPAD at a specified overbias, then initiate laser pulsing and record the time at which the SPAD first registers a pulse. 
Taking \txsim$12,000$ measurements for each SPAD overbias, we extract PDE from the mean interval between photon arrivals divided by the mean measured time-to-first-pulse. 
For the DCR, a time-to-pulse method would be unsuitably slow, so we report the number of dark counts in a certain time interval, thereby potentially slightly overestimating.

\renewcommand{\arraystretch}{1.5}
\begin{table}
\centering

\begin{tabular}{ c | c | c | c | c | c}
              & 422 scatter  & 1092 scatter & Lab & RF  & DCR \\
\hline
Counts (kcps) & 24.22 & 14.29 & 0.84 & 0.16 & 0.11 \\
\hline
Error (stat)  & 0.05  & 0.05  & 0.01 & 0.01 & 0.01 \\
\hline
\end{tabular}
\caption{
\label{tab:countrates}
Various contributions to the background count rate (see main text). 
}
\end{table}

\subsection{Ion-trapping apparatus}

The trap chip is wirebonded first to a ceramic interposer, which is then attached to a UHV compatible circuit board, where $1~\upmu$F capacitors for DC electrodes, SPAD quench resistors, and coaxial connectors for RF and SPAD signals are also located.
The board mounts to an aluminum structure which provides heat-sinking and positions the trap at the center of an octagonal stainless steel vacuum chamber. Neutral Sr is provided by a $3$\s{}mm diameter metal vapor source (AlfaVakuo e.U.) located $6$\s{}cm away from the trap; several aluminum baffles prevent deposition of excess Sr on the trap or interposers. The chamber is pumped with a Nextorr Z combination ion pump/non-evaporable getter.
With a $10$~day bakeout to $120\s^\circ$C, the chamber reaches $<1\times10^{-10}\s$torr with the source off, and $4\times10^{-10}\s$torr with the source at $350\s^\circ$C for loading Sr.
An RF voltage of frequency $38$ MHz and amplitude $70$~V gives rise to radial ion trap frequencies of \txsim$2 \pi \times 5$~MHz, and DC voltages are selected for an axial frequency of $2 \pi \times 700$~kHz.

Detection laser beam paths were optimized to minimize laser scatter into the SPAD, the dominant source of background counts in this experiment; Doppler-cooling and repumping beams, both present during state detection, were parallel to the surface of the chip within 1~mrad.
An achromatic lens was used to focus these beams at the trap center with a Rayleigh range approximately equal to the distance from the trap center to the chip edge, so as to minimize the beam size when in proximity to the leading and trailing edges of the chip.
Detection beam powers were adjusted to optimize detection fidelity; we use $6$\s\muW\s at $422$~nm with a $50$\s\mum\s waist (1/$e^2$ diameter) and $500\s$\muW\s at $1092$~nm (105 \s\mum\s waist). The center of the 1092 beam is displaced 115 \mum\s from the chip surface ($65$~\mum\s from the ion), which reduces the ratio of scatter into the APD to intensity at the ion. While the corresponding intensities are above saturation, detection fidelity is not improved at lower intensity, as we observe increased variability in the count rates for both bright and dark ions.

\subsection{Passive quench circuit and background counts}
\label{sec:cirkit}
The passive quench circuit consists of a $200$~k$\Omega$ quench resistor in series with the SPAD, which has an internal resistance of $25$~k$\Omega$.
An additional $5$~k$\Omega$ probe resistor is placed in series, external to the vacuum chamber.
Cable and in-vacuum chip capacitance are both \txsim$15$~pF, giving rise to the \txsim$500$~ns quench and $3.5$~\mus\s recharge $1/e$ times mentioned in the main text. 
The signal is amplified and a differentiator circuit counts rising edges regardless of where during the recharging time the event occurs. Optimum performance, taking into account speed limitations due to the readout-circuit capacitance, is obtained with bias of $-30~$V for a total count rate $106$~kcps ($38$~kcps) while detecting the bright (dark) state. Table~\ref{tab:countrates} shows contributions to the dark-state count rate; these are dominated by laser scatter, with small additional contributions from stray light in the laboratory, the applied trap RF voltage, and the DCR of the device itself. The contribution from the trap RF voltage is observed to have a time constant of approximately 100 s, suggesting that it may result from heating of the trap by 10-20 K, leading to an  increase in DCR. The device is saturated by about $5\%$ when detecting a dark ion, such that the sum of all contributions to the dark ion rate add to slightly more than 38 kcps.

Given these quench and recharge times, it is instructive to estimate device performance with the use of active quenching electronics, which could potentially be integrated with SPADs like those demonstrated here.
Such circuits achieve sub-microsecond quench and recharge times by detecting the beginning of an avalanche event and arresting it before completion.
Significantly shorter recharge times should enable the device to run at higher overbias without saturation, capitalizing on increased PDE. 
Adaptive readout may enable recharge times on the order of $100$~ns, compatible with operation at $6$~V overbias, which may allow for bright state count rates as high as $150$~kcps with a DCR-limited dark state count rate of $0.6$~kcps.

\subsection{Over-dispersed fluorescence distributions}

The observed dark state count distribution shows an increased width relative to the Poissonian expectation (Fig.~\ref{fig:detect}), with measured variance 20\% larger than a Poissonian distribution of the same mean. 
This may be caused by slow beam pointing drifts of the readout lasers during measurement, affecting scatter seen by the APD. When performing maximum likelihood estimation, we account for the excess variance by modelling the expected photon counts per bin with a negative binomial distribution, a well-known method for modeling over-dispersed Poissonian distributions~\cite{GelmanBayesian2004}. 
We observe good agreement with the expected per-bin photon counts when modeling the count rates with such distributions. Slow laser pointing drifts might be eliminated in future traps through the use of integrated photonics for light delivery.

\end{document}